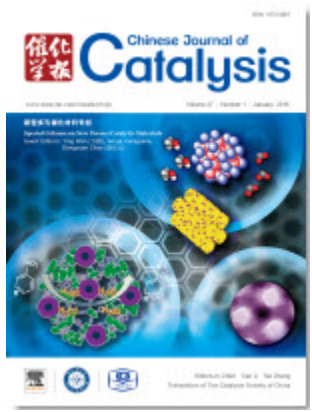

# A facile one-pot hydrothermal synthesis as an efficient method to modulate the potassium content of cryptomelane and its effects on the redox and catalytic properties

# A facile one-pot hydrothermal synthesis as an efficient method to modulate the potassium content of cryptomelane and its effects on the redox and catalytic properties

Huiyan Pan[a], Xiaowei Chen[a, *], Oihane Sanz[b], Miguel A. Cauqui[a], Jose M. Rodríguez-Izquierdo[a], Juan J. Delgado[a, *]

[a] Departamento de Ciencia de los Materiales e Ingeniería Metalúrgica y Química Inorgánica, Facultad de Ciencias, Universidad de Cádiz, Campus Río San Pedro, Puerto Real (Cádiz), E-11510, Spain

[b] Applied Chemistry Department, University of Basque Country (UPV-EHU), Apdo.1072, San Sebastian, 20080, Spain.

*Corresponding authors: juanjose.delgado@uca.es; xiaowei.chen@uca.es

**ABSTRACT**

Cryptomelane has been widely applied as catalyst in oxidation reactions due to its excellent redox properties and low cost. Here, a novel one-pot hydrothermal synthesis using a potassium permanganate aqueous solution as precursor and ethanol as reducing agent has successfully been developed to obtain cryptomelane nano-oxides. This synthetic route makes it possible to control the amount of potassium incorporated into the structure of the cryptomelane by selecting the appropriate synthesis temperature and ethanol initial concentration. Taking advantage of this approach, the effect of potassium concentration on the structural stability and reducibility of the cryptomelane, which are poorly discussed in the literature, has been studied. We have observed that samples with low content of potassium (~11%) show high conversions of CO to $CO_2$ especially at low temperatures. The lower activity of the samples with high K (~16%) contents can be ascribed to the beneficial effect of K on the structural stability of cryptomelane in detriment of labile oxygen on cryptomelane surface.

**Keywords:** Cryptomelane, Pyrolusite, Catalytic CO oxidation, Hydrothermal synthesis, Potassium content.

# 1. Introduction

Homogeneous and heterogeneous catalysts have been predominantly based on the use of precious noble metals such as rhodium, iridium, and platinum [1]. In recent years, due to their scarcity and consequent high cost, increasing efforts have been made in replacing precious metal catalysts with more abundant transition metals [2,3]. In this context, manganese oxides are emerging as appealing alternatives to noble metals due to their abundance in the Earth and the singular reactivity of manganese that exhibits a wide variety of oxidation states, being the most common +2, +3, +4, +6, and +7.

Manganese oxides octahedral molecular sieves (OMS) are very attractive materials in different fields such as catalysis, adsorption-separation, ion-exchange and electrochemistry because their channel structure can host different cations [4–7]. OMS-2 are formed by double chains of edge-sharing Mn(IV)$O_6$ octahedral which are linked forming $2 \times 2$ and $1 \times 1$ tunnel structures (Figure S1) [8]. Normally, the small $1 \times 1$ channels (2.3 Å) are empty, while water or different stabilizing cations are commonly located in the large $2 \times 2$ channels (4.6 Å). In the literature, we can find OMS-2 samples that consist in stabilizing metal ions such as $Ba^{2+}$ (hollandite), $K^+$ (cryptomelane), $Na^+$ (manjiroite) and $Pb^{2+}$ (coronadite). Potassium based OMS-2 (cryptomelane) has received a great deal of attention in catalysis and it has been used not only as supports but also as catalysts [9–12]. Cryptomelane has exhibited excellent activity in different oxidation reactions of interest in environmental protection, such as CO oxidation [10,13–15], decomposition of methylene blue [16], $NO_x$ removal [17], VOCs oxidation [18] or total oxidation of small alkanes [19]. Among these reactions, CO oxidation has been widely studied as a reference catalytic test in the fields of surface science and heterogeneous catalysis due to its technological importance and fundamental interest [20].

Several research studies indicate that the high activity of cryptomelane can be ascribed to their open tunnel structure, the presence of mixed-valence manganese ions ($Mn^{4+}$ and $Mn^{3+}$) and, what is even more important, the presence of labile oxygen [13,21,22]. However, some fundamental aspects to understand the origin of the labile oxygen still need to be addressed. This knowledge would be crucial to optimize the redox properties of pure cryptomelane and improve its catalytic activity. In this sense, it is worth to mention that in order to accommodate the potassium in the center of the 2 × 2 tunnels, manganese atoms of the structure are partially reduced to keep the charge balance. Therefore, the concentration of potassium in the cryptomelane structure may play an important role in its redox properties. It is thus quite surprising that very little attention has been paid to the effect of the potassium content on the reducibility and stability of cryptomelane. In fact, it must be highlighted that the exact values for potassium contents are not commonly provided in the literature and cryptomelane-type materials are frequently denoted as $K_xMn_8O_{16}$. In spite of this, a few recent results suggested an important role for $K^+$ in the catalytic performance of these materials. Hou et al. stated that the catalytic removal of benzene could be significantly enhanced by increasing the $K^+$ content [23]. The authors reported some interesting theoretical calculations demonstrating that potassium may strongly affect the reducibility of the samples. Even more recent results revealed that potassium concentration in the tunnels increases the content of oxygen vacancy and it is beneficial for ozone elimination [24] and dimethyl ether combustion [25]. In both cases, the authors used a post-processing method with KOH to increase the potassium concentration of previously synthesized α-$MnO_2$ and cryptomelane nanoparticles. It is obvious, therefore, the interest of a deeper understanding of the effect of potassium on the redox properties and structure stability of cryptomelane. On the other hand, the development of novel synthetic routes to obtain, preferentially by a single-step, cryptomelane with the optimized potassium concentration is of great importance.

In the present work, a simple and scalable hydrothermal method using ethanol to reduce a $KMnO_4$ aqueous solution has been developed to prepare cryptomelane-type manganese oxides. The effects of the ethanol concentration and reaction temperature on the crystallinity, textural properties and morphology have been systematically investigated. This synthetic route allowed us to modulate the $K^+$ concentration of the samples, making possible an in-depth study of its effect on the redox properties and the stability. The catalytic performance of these low-cost materials for the oxidation of CO reaction has been evaluated and correlated with the results obtained from the physicochemical, textural and structural characterization of the samples.

## 2. Experimental

### 2.1 Materials and catalyst preparation

The synthesis was carried out on a custom-built six-port autoclave and each Teflon vessel has a volume of 150 mL. The autoclave was placed inside of a stove equipped with a rotation device that was fixed at 50 rpm. A picture of the experimental device is included in the supporting information (Figure S2). The synthetic process briefly consists as following: 50 mL of a potassium permanganate (> 99%, Scharlau) aqueous solution (10 g/L) was introduced into the Teflon liner vessels and subsequently absolute ethanol (99.98%, VWR) was dropwisely added. The concentrations of ethanol used in the synthesis medium were 50, 170, 960 and 1550 mM. After sealing the autoclave, the temperature was raised up to 80, 100 or 120 ºC for 12 h. After cooling down to ambient temperature, the weight of each vessel was measured and compared with the initial weight to ratify the integrity of the vessels during the synthesis. The samples were filtered and washed once with 50 mL of miliQ water. Then, they were dried at 70 ºC overnight and further calcined at 350 ºC for 6 h in a muffle oven. Finally, the obtained solid samples were ground and sieved with a 100 mesh sieve. It must be pointed out that when ethanol was not added, the solution remained purple after the hydrothermal treatment and no solid was obtained. On the other side, the manganese concentration remaining in the reaction medium after the synthesis was below 2 mg/L, reaching a very high yield of 99.9%.

In order to facilitate the sample identification, we coded them using the synthesis temperature and the initial ethanol concentration (mM) in the reaction medium. For instance, the manganese oxides prepared at 100 ºC with an initial ethanol concentration of 1550 mM is coded as 100-1550. Besides, a standard cryptomelane sample K-OMS-2 was prepared by refluxing method as it was previously described in Ref [11]. This sample is labeled as K-OMS-2 in the text.

*2.2 Physical, chemical and structural characterization*

X-ray diffraction (XRD) patterns were collected using a Bruker diffractometer AXS, model D8 Advance, employing Mo $K_{\alpha1}$ radiation source (0.709319 Å) and operating at 50 kV and 50 mA. The scan range was from 3º to 38º with a step size of 0.02º and a counting time of 2 s per step. The K and Mn contents of the catalysts were determined by X-ray fluorescence (XRF) technique using a M4 TORNADO (BRUKER) instrument. The composition was obtained by calculating the average of one hundred measurements performed with a 25 mm spot. The Brunauer-Emmet-Teller (BET) specific surface areas of the catalysts were measured by nitrogen adsorption at −196 °C on an autosorb iQ$_3$ instrument from Quantachrome Instruments.

X-ray photoelectron spectroscopy (XPS) experiments were carried out in a Kratos Axis Ultra DLD equipped with a monochromatized Al *Kα* X-ray source (1486.6 eV) operating with an accelerating voltage of 15 kV and 10 mA current. Spectra were acquired in the constant analyzer energy mode, with a pass energy of 20 eV. Powder samples were pressed into pellets and stuck on a double-sided adhesive conducting tape. The samples were analyzed without any pretreatment. Surface charging effects were compensated by using the Kratos coaxial neutralization system. Charging effects were corrected by adjusting the binding energy of the C (1*s*) peak at 284.8 eV. CasaXPS software (version 2.3.19) was used for the data analysis.

Scanning Electron Microscopy (SEM) images were taken on a FEI Nova 450 electron microscope with an accelerating voltage of 3 kV. Catalysts were also analyzed by transmission electron microscopy (TEM) using a JEOL2010F instrument equipped with an Oxford INCA Energy 2000 Energy dispersive X-ray spectrometer (XEDS). This instrument has a spatial resolution at Scherzer defocus conditions of 0.19 nm. High Angle Annular Dark Field-Scanning Transmission Electron Microscopy (HAADF-STEM) technique using an electron probe of 0.5 nm of diameter was used for recording the XEDS.

Temperature-programmed desorption (TPD) and CO-temperature-programmed reduction (CO-TPR) were carried out with a U-type quartz reactor and 75 mg of the sample. A mass spectrometer (Thermostar GSD301T1, Pfeiffer Vacuum) was coupled to the equipment in order to register the product evolution and the experiment temperature. All the thermal treatments were performed with a heating rate of 10 °C/min and a flow of 60 mL/min. In order to eliminate adsorbed water and carbonates, the samples were firstly heated up to 350 °C in a 5% $O_2$/He atmosphere and the temperature was maintained for 1 h. After cooling down to room temperature, pure He was introduced into the reactor and kept flowing for 1 h. After the cleaning treatment, the TPD experiments were performed up to 900 °C in He. On the other hand, 5% CO/He was switched into the reactor and kept 0.5 h at room temperature before CO-TPR. After that, the reactor was heated up to 950 °C.

*2.3 Catalytic activity for CO oxidation*

The catalytic assays for CO oxidation over manganese oxide samples were carried out in a U-type quartz reactor and using 25 mg of catalyst diluted by 50 mg of SiC. Initially, the sample was pretreated in a 5% $O_2$/He flow at 350 °C for 1 h. After cooling down to room temperature, pure He was purged for 0.5 h. The reaction mixture (1% of CO, 0.6% of $O_2$ and 98.4% of He with a total flow rate of 100 mL/min) was switched to the reactor and maintained 0.5 h at ambient temperature. Then, the reactor was heated to a maximum temperature of 400 °C with a heating rate of 10 °C/min. The reactant and product evolution was obtained by analyzing the stream with a Pfeiffer Vacuum Thermostar mass spectrometer (model GSD301T1).

# 3. Results and discussion

*3.1 Structural and textural analysis*

The obtained manganese oxides were firstly studied by XRD to determine their crystalline structures. Figure 1 shows the XRD patterns of all samples and the K-OMS-2 obtained by a reflux method. We can observe that the diffractograms of the samples synthesized at 80 and 100 °C can be indexed to pure tetragonal cryptomelane (PDF 44-1386) and no additional diffraction peaks are observed. However, in the diffractograms of the samples, the intensities of peaks are strongly influenced by the ethanol concentrations. The samples obtained with lower concentration of ethanol exhibited narrow and well-defined diffraction peaks, while those prepared with higher ethanol contents displayed broad peaks and therefore lower crystallinity. Operating the hydrothermal reaction at 120 ºC, cryptomelane is obtained using lower ethanol concentrations (50 and 170 mM), but samples synthesized with higher ethanol concentrations mainly contain pyrolusite phase (PDF 24-0735). These results indicate that cryptomelane can be obtained in a wide range of the synthesis conditions and that crystallinity of the sample can be modulated by the temperature and ethanol concentration. Indeed, adding low ethanol contents improves the crystallinity of cryptomelane-type manganese oxide samples.

Table 1 includes the BET specific surface area ($A_{BET}$) of each sample and the $A_{BET}$ as a function of ethanol concentrations is displayed in Figure 2 (top). The $N_2$-physisorption isotherms are included in Figure S3. The pyrolusite-containing samples obtained at 120 ºC and high ethanol concentration exhibited the smallest specific surface areas (50 and 13 $m^2/g$). On the other hand, the BET surface area of the cryptomelane samples varies from 100 up to 200 $m^2/g$. The cryptomelane samples with low specific surface areas (100–135 $m^2/g$) exhibit high crystallinity, what can be ascribed to the presence of large particles. In addition, the cryptomelane samples with broad diffraction peaks have high BET specific surface areas (160–200 $m^2/g$) and, thus, smaller particle size.

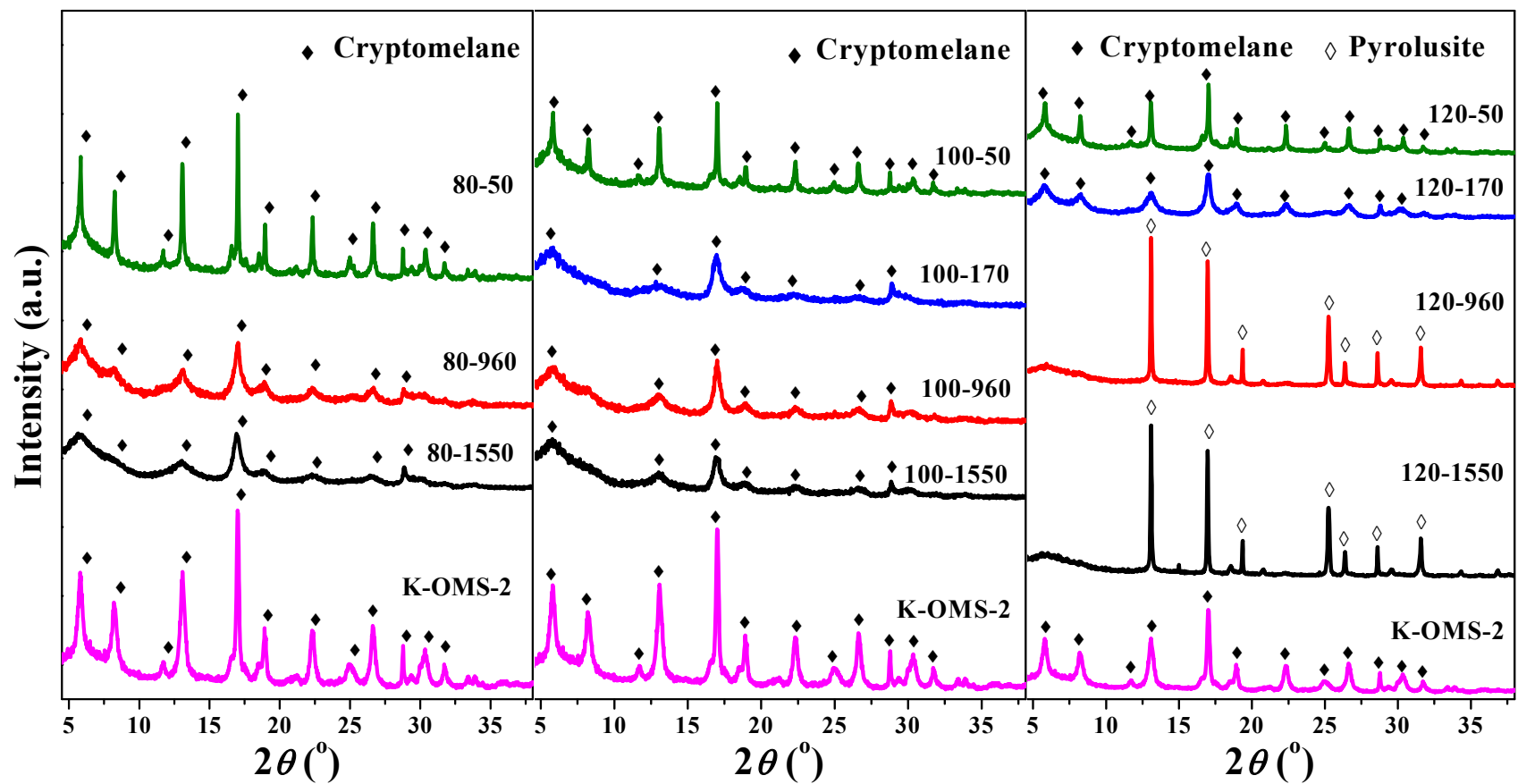

**Figure 1.** XRD patterns of the samples obtained at 80 (left), 100 (middle) and 120 ºC (right). K-OMS-2 is included as reference. The main peaks are assigned to cryptomelane (♦) or pyrolusite (◊).

**Table 1.** Composition, specific surface area and CO oxidation activity of all manganese oxide samples.

| Sample | K (mol.%) in K + Mn[a] | $A_{BET}$ (m²/g) | CO conversion at 75 °C (%) | Rate at 75 °C (µmol·m$^{-2}$·g$^{-1}$·min$^{-1}$) | $T_{50}$[b] (°C) | $T_{90}$[b] (°C) | Group |
|---|---|---|---|---|---|---|---|
| 80-50 | 16.4 | 106 | 0.0 | 0.00 | 195 | 268 | A |
| 80-960 | 11.5 | 193 | 23.8 | 2.20 | 102 | 157 | B |
| 80-1550 | 11.1 | 189 | 18.4 | 1.74 | 108 | 146 | B |
| 100-50 | 16.4 | 127 | 0.0 | 0.00 | 171 | 204 | A |
| 100-170 | 12.6 | 189 | 16.2 | 1.53 | 109 | 158 | B |
| 100-960 | 10.9 | 176 | 19.0 | 1.93 | 111 | 144 | B |
| 100-1550 | 10.6 | 165 | 17.2 | 1.86 | 110 | 143 | B |
| 120-50 | 16.5 | 135 | 1.0 | 0.13 | 172 | 224 | A |
| 120-170 | 10.2 | 162 | 23.7 | 2.61 | 112 | 194 | B |
| 120-960 | 4.7 | 50 | 2.9 | 1.04 | 191 | 248 | C |
| 120-1550 | 0.7 | 13 | 0.0 | 0 | 215 | 268 | C |
| OMS-2 | 12.1 | 70 | 5.0 | 1.27 | 155 | 192 | -- |

a: determined by XRF.

b: the temperature of achieving 50% or 90% of CO conversion.

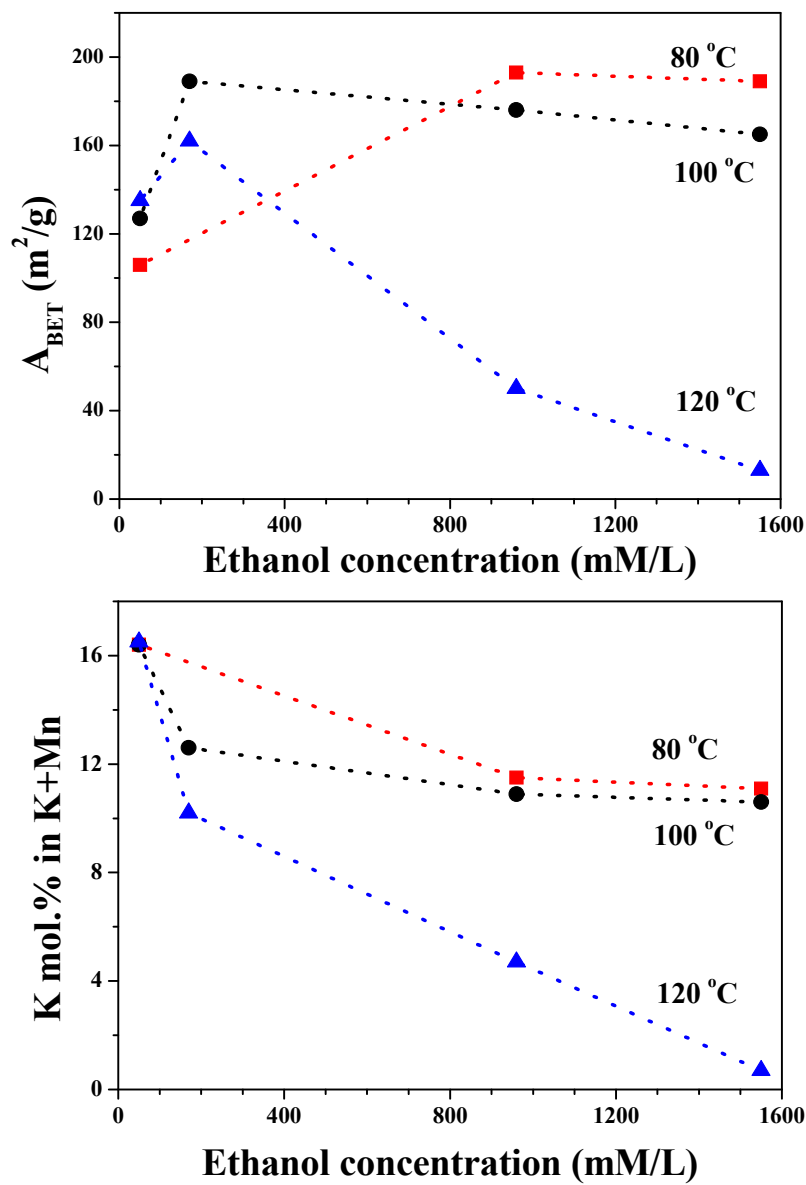

**Figure 2**. Specific surface area (top) and potassium content (bottom) of the obtained samples as a function of initial concentration of ethanol.

*3.2 Effect of the synthesis conditions on the K$^+$ concentration*

As it was mentioned in the introduction, the amount of potassium in the 2 × 2 tunnels will determine the amount of Mn$^{3+}$ generated for charge balancing. Furthermore, the potassium content of cryptomelane may affect its redox properties, structure stability and the catalytic performance. The potassium contents of the samples are included in Table 1 and Figure 2 (bottom). We observe that at 80 and 100 ºC, the potassium molar ratio decreases from 16.4% to 10.6% by increasing the initial ethanol concentration from 50 to 1550 mM. Nevertheless, the X-ray diffraction results indicate that all the samples have the cryptomelane structure. Obviously, it is possible to obtain cryptomelane with different potassium content by adjusting the synthesis conditions. In addition, the diffraction peaks do not shift, demonstrating that the incorporation of the potassium to the channel structure doesn't affect the cell parameters. The latter result is in good concordance with those reported in the literature [23,26]. Our results suggest that samples with high potassium contents (16%) exhibit an apparently higher crystallinity and lower $A_{BET}$ than the nano-oxides with lower potassium contents. These results may also imply that the stabilizing effect of $K^+$ promotes the formation of bigger well-defined nano-particles and, consequently, the specific surface area decreases.

The K content of the samples obtained at the 120 ºC dramatically decreases from 16.5% to 0.7% by increasing the ethanol concentration. This result is directly related with the formation of pyrolusite when an ethanol concentration higher than 170 mM is used. Under these conditions, the potassium permanganate is mainly transformed into pure pyrolusite. The potassium can't be incorporated to the pyrolusite structure and therefore remained in the form of $K^+$ in the aqueous solution, which can be washed away from the mother solution by water [27]. Even though the XRD results previously discussed indicate that the samples obtained at 120 ºC are apparently pure cryptomelane or pyrolusite, the TEM studies reveal that the samples synthesized with an initial ethanol concentration higher than 170 mM are mixtures of both kinds of nanoparticles. The absence of cryptomelane phase in the XRD pattern of 120-960 and 120-1550 is ascribed to the low content of cryptomelane and its small particles size. These results will be shown below in the microscopy study part. It is also remarkable that the manganese oxide samples obtained at three different synthesis temperatures with an ethanol concentration of 50 mM have nearly the same K content and good crystallinity.

We can conclude that the potassium content of the sample can be easily tailored by the synthesis conditions. In general, high concentration of ethanol and high synthesis temperature reduce the potassium content. The effect of the potassium content on the redox and catalytic properties will be discussed later. In order to facilitate the discussion and reduce the data redundancy, due to the large number of samples obtained, they were divided into three types depending on their physicochemical properties. The first group (A) of samples is essentially cryptomelane with high crystallinity, high specific surface areas (100–135 m$^2$/g) and a potassium content around 16.4%. The second group (B) of samples is mainly cryptomelane with low crystallinity, even higher A$_{BET}$ (160–200 m$^2$/g) and a potassium content between 10 and 12.6 %. Finally, the third group (C) is mainly pyrolusite with low surface areas (50–13 m$^2$/g) and low potassium content (4.7–0.7%). The samples are classified accordingly with this criterion in Table 1. It must be mentioned that, as it will be discussed later, samples of the same group also have similar microstructure, redox properties and catalytic activity.

### *3.3 Morphology, microstructure and composition analysis by electron microscopy*

SEM, HRTEM and HAADF-STEM techniques have been used to characterize the samples at the nanometric scale. Figure 3 includes representative SEM and TEM images of each previously described groups. Additional images are presented in Figures S4, S5 and S6. According to the SEM images, short nanorods with a length of 150–400 nm and a thickness of 30–90 nm are observed in samples of Group A. The HRTEM images of the sample can be interpreted as cryptomelane (Figure 3B), which is in good concordance with the X-ray diffraction results. The SEM and TEM images of samples included in Group B are mainly formed by nanorods with smaller thickness (5–15 nm) and a length that can vary from 10 to 2000 nm (Figure 3C and S4). The analysis of HRTEM and HAADF images (Figure S4) indicate that long and apparently thick rods are really bundles of much more thinner (5–15 nm) nanowires. The in-depth analysis of HRTEM images demonstrates that both long and short rods have cryptomelane structure. Figures 3D and S4 demonstrate that the channel structure is well defined even in the short (5–7 nm) rods. Additionally, HAADF-STEM, in combination with XEDS, was used to study the chemical composition of individual nanoparticles. Figure S4 shows that only small vibrations of the K/(Mn + K) ratio, between 10.9 and 13.5 %, are observed and that small and large particles have similar composition. The HRTEM images and the DDP analysis of Groups A and B reveal well defined nanoparticles with good crystallinity, although the XRD patterns of Group B show broader peaks with lower intensity than those observed for Group A. Therefore, the weak diffraction peaks obtained for samples of Group B can be attributed to their smaller size than the nanoparticles of Group A.

Finally, samples included in Group C are mainly formed by long and very thick nanorods (Figure 3G), which is in good accordance with the low BET surface area of these samples and the high crystallinity observed in the X-ray diffractograms. In addition, the Digital Diffraction Pattern (DDP) of those rods can be indexed to $\beta$-$MnO_2$, being in good agreement with the XRD results (Figure 3H). However, small rods can also be observed in sample 120-1550. The chemical analysis of large and small rods was conducted by XEDS analysis (Figure S5). Regarding the cation composition, both potassium and manganese can be found in the small nanorods, while manganese was the only cation detected in the large rods. In fact, HRTEM images included in Figure S5 clearly shows a spacing of 6.9 Å, which corresponds to the channel periodicity of cryptomelane. Considering their small size and molar potassium content (10–12%), we can conclude that these nanoparticles are very similar to those obtained in Group B. The results also show that the amount of short rods in sample 120-1550 is low, but in the case of sample 120-960 we can observe that the big rods are completely coated by them (Figure S6). It must be pointed out that the XRD results reveal that both samples are $\beta$-$MnO_2$ and no additional peaks were observed. This can be explained if we consider that the $K_xMn_8O_{16}$ rods are similar to those obtained in all the samples of Group B and that their diffraction patterns are poor. The HRTEM images of this phase obtained in all the samples show a good crystallinity and the low intensity and broad peaks in the XRD can be related to the small thickness of the particles. It is obvious that a higher concentration of the cryptomelane nanocrystals in the sample would increase the potassium content of the sample and the BET specific surface area. Therefore, this result also explains why the BET specific surface area and the potassium content (Table 1) are much higher in the case of sample 120-960 than that in the sample of 120-1550. It is evident that depending on the synthesis conditions, the obtained sample can have a mixture of all the phases previously described, but they have been included in the most appropriate group according to their chemical properties and structure for an appropriate discussion of the results.

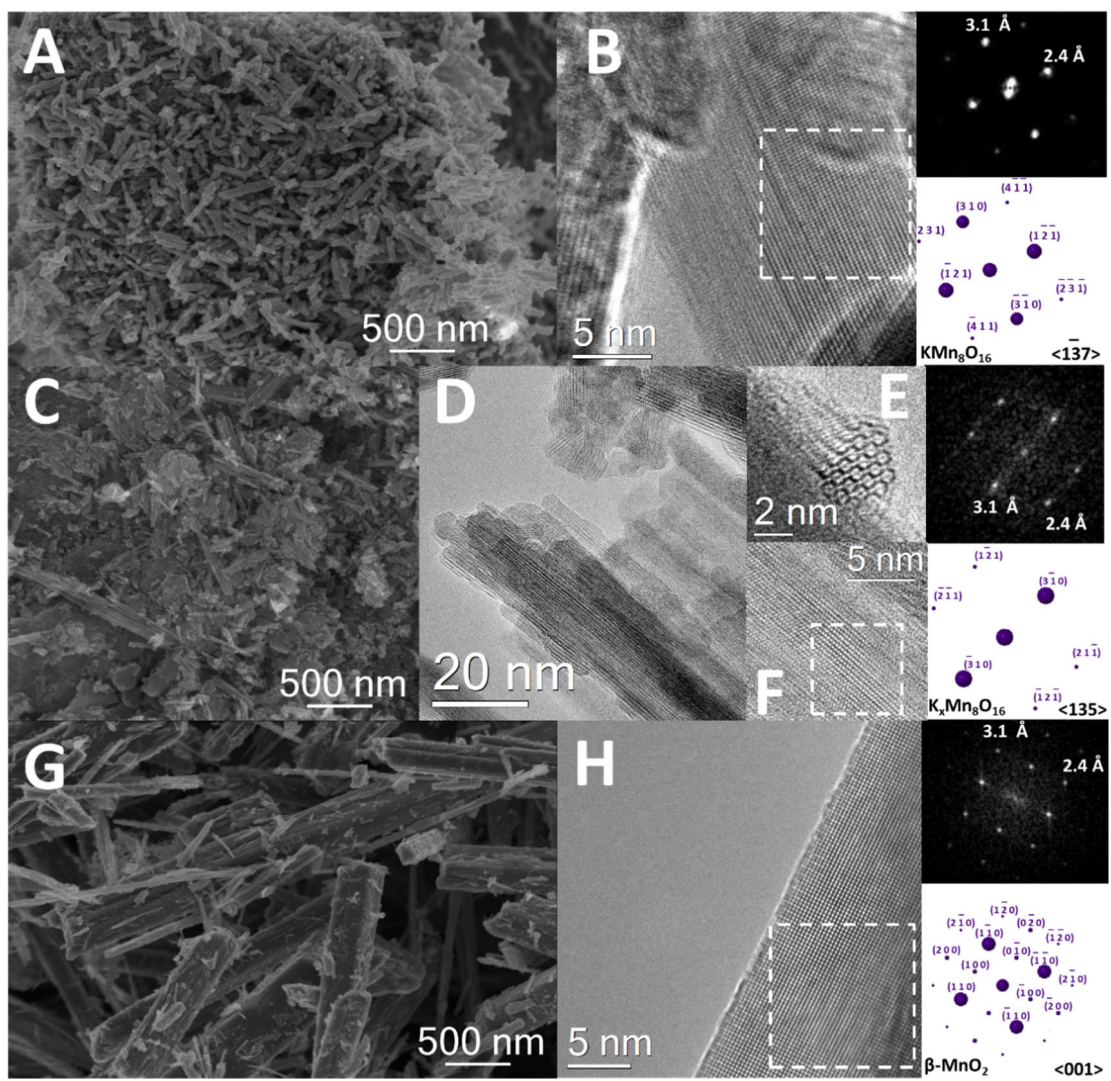

**Figure 3.** SEM and TEM images of samples 100-50 (A, B), 100-1550 (C, D, E, F) and 120-1550 (G, H). The DDP of the selected area and its analysis is also included on the right. Samples 100-50, 100-1550 and 120-1550 are representative of Groups A, B and C, respectively.

*3.4 Manganese oxidation state, reducibility and thermal stability*

XPS analysis was used to determine, on the surface, the oxidation state of manganese cations, the concentration of potassium and adsorbed oxygen. The results are shown in Figure 4 and Table S1. The average oxidation state (AOS) of manganese was estimated by analyzing the Mn 3*s* region (Figure 4, left). According to the literature, the AOS can be obtained by calculating the binding energy difference ($\Delta E_{3s}$) between the two peaks observed in the Mn 3*s* region and applying the following formula: AOS = 8.956 – 1.126·$\Delta E_{3s}$ [6,28]. Samples of Group C have an AOS of manganese close to 4, what is in good accordance with the previously discussed results indicating that this sample is mainly β-MnO$_2$. On the other hand, the obtained AOS of Groups A and B were 3.73 and 3.81, respectively. XPS results also suggest that the concentration of K on the surface of Group A (27.2%) is higher than that in Group B (17.9%). Both results corroborate that the potassium incorporated in the channel structure of cryptomelane reduces the oxidation states of manganese to keep the charge balance. Interestingly, the O/Mn ratio is close to 2 in all the cases, what means that cryptomelane keeps the stoichiometry of α-MnO$_2$. This result also indicates that the oxygen content is not affected by the potassium concentration. It is also of interest to underline that the content of superficial potassium obtained by XPS is bigger than the bulk concentration of potassium obtained by XRF. In other words, the surface is enriched by potassium. In fact, the Group A has a high potassium content of 27.2% and therefore a higher amount of Mn$^{3+}$ in concordance with the AOS. This result may also imply that there is a higher concentration of Mn$^{4+}$ in the center of the cryptomelane nanoparticles in Group B.

O 1*s* spectra of the samples are displayed in Figure 4 where the O 1*s* spectral region is deconvoluted to two different oxygen species: (1) Mn-O-Mn lattice oxygen at 529.5–529.7 eV, and (2) superficial hydroxyl groups or adsorbed oxygen at 531.0–531.5 eV [6,26,29]. The data analysis results are listed in Table S1. The contribution of the commonly denoted as adsorbed oxygen is higher in Group B (38.2%) than that in other two groups, hinting a higher concentration of labile oxygen that may play an important role in the catalytic activity. The results can be attributed to the higher surface of the samples in Group B and the higher amount of Mn$^{4+}$ that can be inferred from its higher AOS.

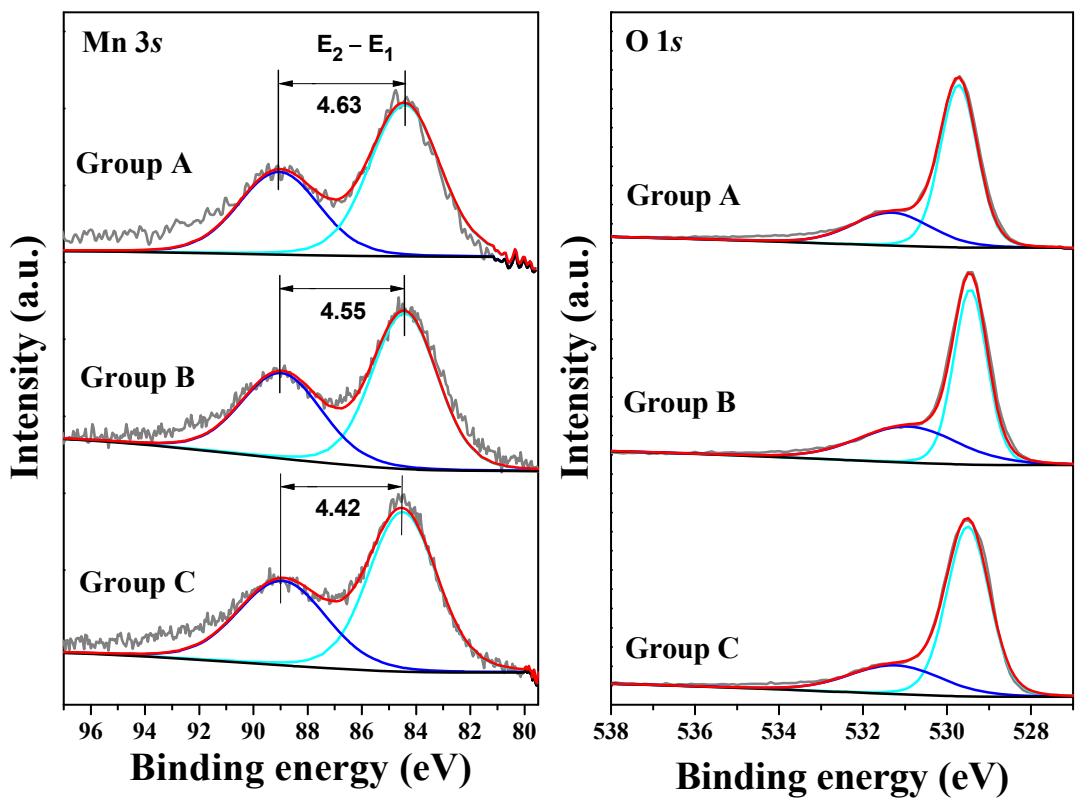

**Figure 4.** XPS spectra of Mn 3*s* (left) and O 1*s* (right) of representative samples from Group A (100-50), Group B (100-1550), and Group C (120-1550).

The effect of the catalyst reducibility on the catalytic activity in oxidation reactions has been intensively studied and it is commonly considered as one of the most influential properties. Manganese oxides, including cryptomelane, are nonstoichiometric oxides and a mixture of $Mn^{3+}$ and $Mn^{4+}$ can be found on the surface of the sample, which is assumed that has a beneficiary effect on the catalytic activity [12,21]. TPD and CO-TPR experiments were conducted to evaluate the reducibility of the prepared samples and try to identify different oxygen species involved in the catalytic oxidation of CO. The $O_2$ release of the samples was evaluated by recording the m/z 32 using a mass spectrometer during TPD process (Figure 5). We can observe that each group exhibits dramatically different oxygen release, but only slight difference in the same group can be found. The changes inside of the same group only moderately affect the position of the peak and its relative intensity. It is clear that Group B starts to release oxygen at lower temperatures (300 ºC) and the $O_2$ profiles show three main peaks centered at around 418, 540 and 745 ºC. Similar results have been reported in the literature [21,30] for cryptomelane samples, although the area related to the first peak is bigger for the samples of Group B obtained in this study. The peak observed at low temperature is commonly attributed to desorption of surface oxygen species and/or the presence of labile surface lattice oxygen species normally associated to $Mn^{4+}$ [21,30]. According to the literature, the second peak is mainly assigned to the transformation of cryptomelane to $Mn_2O_3$ and its subsequent reduction to $Mn_3O_4$ is the origin of the third peak [21].

Even though both Group A and B, as shown in XRD results, are cryptomelane-type manganese oxide, their TPD profiles are quite different and only the peak at high temperature is observed for samples included in Group A. According to the literature, the missing peaks at 418 and 540 $^o$C are related to the reduction of $Mn^{4+}$ and the only observed peak corresponding to the reduction of $Mn_2O_3$ to $Mn_3O_4$ for Group A moves to higher temperature (795 $^o$C) than that for Group B. Samples of Group A have a higher potassium content and then the amount of Mn(IV) is smaller as it was corroborated by XPS. However, hypothetically it would need a K atomic ratio of 50% to fully reduce all the Mn(IV) and the highest content was much smaller (16%). Additionally, the intensity of the unique observed peak is much higher than the peak obtained for Group B at a similar temperature. To the best of our knowledge, Sun [25] and Zhu [24] are the only authors that reported similar results. Both authors observed that the contribution of the peak at low temperature was significantly smaller after treating cryptomelane samples with KOH. Therefore, both syntheses were carried out with a high potassium content and it would be reasonable that operating at those conditions, as it happens in our case, the amount of potassium incorporated into the channels would be higher. We propose that the higher amount of potassium in the channels would enhance the stability of the nanoparticles and they would "directly" decompose to $Mn_3O_4$. In fact, as it was mentioned before, the area of the peak associated to this transformation is very similar to the area of the last two peaks exhibited by samples of Group B. It can be concluded that the cryptomelane samples present in Group A are much more stable than those samples in Group B. Hence, we can consider that the higher amount of potassium stabilizes the cryptomelane structure. However, we must mention that the results obtained by Hou et al. using cryptomelane samples with low potassium content (< 8%) suggested that the reducibility of cryptomelane was enhanced by increasing the potassium content [23]. They also included some theoretical calculations indicating that the higher amount of potassium, the lower the removal energy of lattice oxygen. However, their studies were performed with potassium contents lower than 8%. Consequently, their conclusions are complementary to ours and it is reasonable to assume that the incorporation of potassium to the channel structure of α-$MnO_2$ improves its reducibility, but at higher contents the structure is more stable and the reducibility at low temperature is worse.

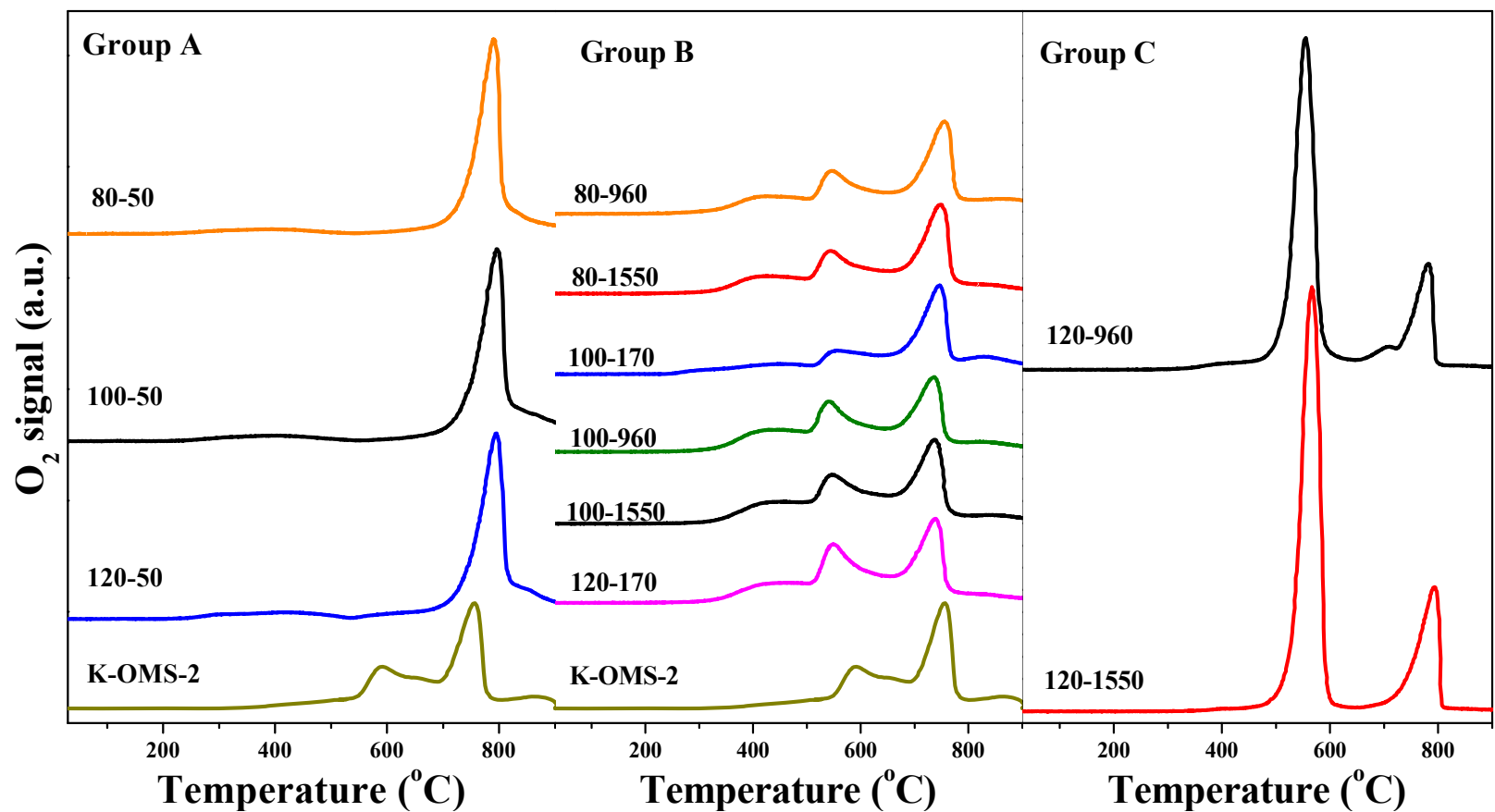

**Figure 5** O$_2$ release curves during TPD experiments of samples included in Group A (left), B (middle) and C (right).

Regarding Group C, two main $O_2$ release peaks can be observed at around 560 and 785 ºC. As revealed by XRD results, Group C is mainly formed by pyrolusite-type manganese oxide. Usually, pyrolusite reduction under inert gas takes place in two steps related with the phase transformation from $MnO_2$ to $Mn_2O_3$ and from $Mn_2O_3$ to $Mn_3O_4$ [31]. It must be mentioned that the ratio between the areas of the observed first and the second peak is 73:25 in the case of the sample 120-1550 and 70:30 in the case of the sample 120-960. In both cases, it is really close to the theoretical ratio (75:25) of the previously described reduction mechanism. Finally, it is of interest to comment that sample 120-960 also exhibits a small peak prior the reduction peak associated to $Mn_3O_4$. This peak can be related to the presence of the small cryptomelane nanorods that were observed by SEM.

During the TPD experiment, m/z 18 (related with water) was also recorded and the results obtained for the groups mainly formed by cryptomelane (Groups A and B) are shown in Figure S7. We can observe that all samples release water at high temperature, although the amount is much higher in the samples of Group B. We must remember that $H_3O^+$ and water can be incorporated in the channels of the α-$MnO_2$ to stabilize its structure. Gao et al. [32] and Espinal et al. [33] have carefully studied the structure of cryptomelane-type manganese oxides and they have estimated that its formula is $K_{0.11}(H_3O)_{0.05-0.08}MnO_2$. According to their results, the K/(K + Mn) ratios are around 9% and if we consider the $H_3O^+$, as $(H_3O^+ + K)/(K + H_3O^+ + Mn)$, the total atomic ratio of stabilizing cation is around 15%. Comparing these results with ours included in Table 1, we can conclude that samples of Group A, have an amount of K (16%) close to the maximum atomic ratio of stabilizing cations that the cryptomelane can accept and therefore less water/$H_3O^+$ can be found in the channels. So under the synthesis condition of Group A, it is promoted the incorporation of the potassium into the channel structure of the α-$MnO_2$ and therefore the maximum amount of water that can be introduced is much smaller. This information supports the assumption that $K^+$ is better than water/$H_3O^+$ as stabilizing cation and that the stability of the cryptomelane structure increases by increasing the potassium content.

In order to elucidate the reducibility of the sample in the presence of CO, one sample of each group and the reference sample (K-OMS-2) were also studied by CO-TPR. Figure 6 displays the CO (m/z = 28) consumption and $CO_2$ (m/z = 44) desorption curves. Firstly, it is clear that CO consumption is coupled to the $CO_2$ desorption in all the cases. The reference sample (K-OMS-2) shows three $CO_2$ desorption peaks at 288, 327 and 410 °C. These results are in good agreement with previously reported results [34,35]. The first reduction peak of cryptomelane-type manganese is owing to the remove of labile oxygen, while the second reduction peak can be attributed to the reduction of $K_xMn_8O_{16}$ to $Mn_3O_4$ and the third reduction peak is due to the reduction of $Mn_3O_4$ to MnO [34]. Interestingly, Figure 6 clearly shows that even though the CO consumption profiles of Group A and B are similar to the reference cryptomelane sample, they can be reduced even at room temperature in the presence of CO. However, the reduction of the sample is more obvious at low temperature for Group B and the positions of the main peaks shift to higher temperature in the case of Group A. Moreover, the shoulder at low temperature reveals the presence of labile oxygen species with low Mn-O strengths. In the case of Group B, this is in good agreement with the TPD results, but Group A does not show a good reducibility under inert atmosphere at low temperature. It is obvious, in this case, that the interaction of the sample with CO favors the reduction of the samples in Group B. It should be pointed out that CO is one of the reactants of our catalytic tests and that therefore the results are expected to be of great interest to rationalize the catalytic activity results. Finally, $CO_2$ desorption of 120-1550 (Group C) displays only two peaks at 330 and 374 °C, both are coupled to the CO consumption. The first peak should be corresponding to the reduction from $MnO_2$ to $Mn_3O_4$, while the second peak is ascribed to the transformation from $Mn_3O_4$ to MnO.

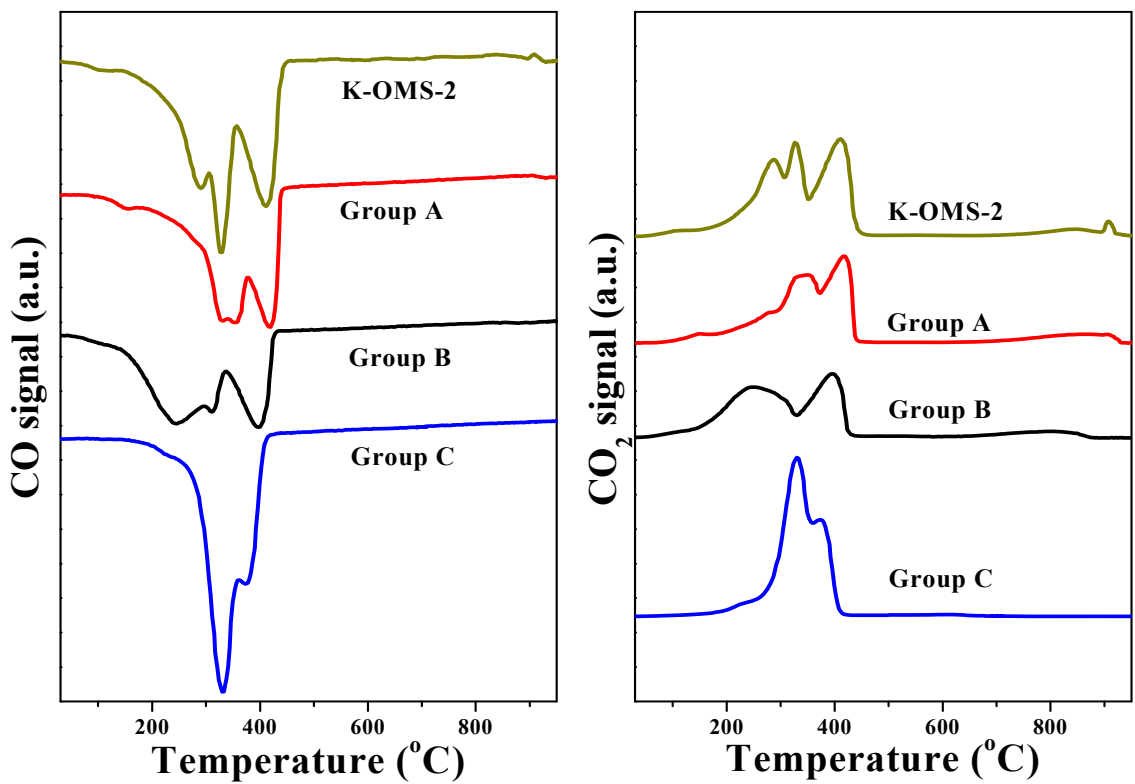

**Figure 6.** CO consumption (left) and $CO_2$ production (right) obtained during CO-TPR of the representative samples from Group A (100-50), Group B (100-1550), Group C (120-1550) and K-OMS-2.

*3.5 Catalytic performance for CO oxidation*

The prepared manganese oxides were finally tested in CO oxidation to check their catalytic properties. In order to facilitate the comparison and the discussion, Figure 7 includes the results obtained for the reference sample (K-OMS-2) and one manganese oxide sample of each group. Moreover, the results of all the catalytic tests are included in Figure S8. Additionally, the temperature necessary to reach 50% and 90% conversion of CO ($T_{50}$ and $T_{90}$) for each sample is displayed in Table 1. Firstly, it is remarkable that the total CO conversion of the most active samples (Group B) is achieved below 200 °C. In fact, samples of Group B are much more active than the reference cryptomelane sample obtained by a reflux method. For this group, the obtained values of $T_{50}$ varied between 102 and 112 °C. In the case of Group A, the achieved $T_{50}$ were around 70 °C higher (170 °C) than that for Group B. Finally, in the case of Group C, the $T_{50}$ reached 191–215 °C, exhibiting a much poor catalytic properties than other two groups. Since Group B is the most active, it seems that higher surface areas and a K/(K + Mn) ratio around 11% leads to higher catalytic activity for CO oxidation.

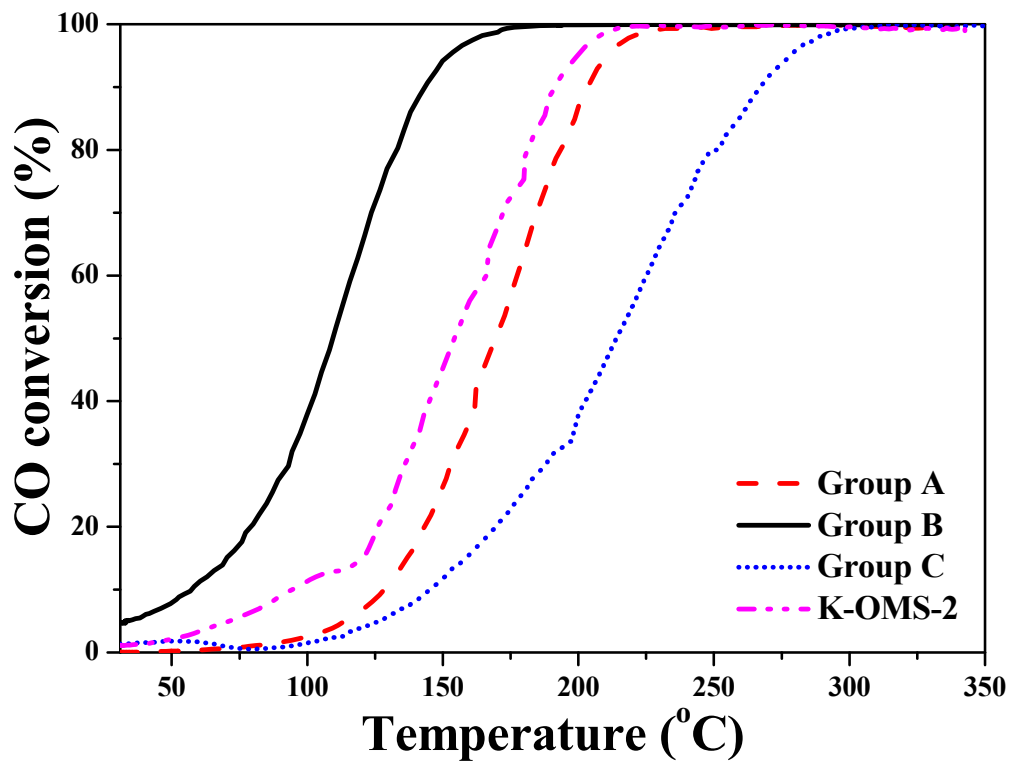

**Figure 7.** CO conversion over representative catalysts from Group A (100-50), Group B (100-1550), Group C (120-1550) and K-OMS-2.

In order to compare the catalytic activity more reasonably, the conversion and reaction rate at 75 °C of all samples are listed in Table 1 for quantitative analysis. At 75 °C, there is nearly no activity for cryptomelane samples in Group A and thus the reaction rate is nearly zero. On the other hand, the cryptomelane-type samples in Group B are very active for CO oxidation. Especially the catalysts 80-960 and 120-170 show conversions almost 5 times higher than K-OMS-2 at 75 °C. Besides, the reaction rate normalized by the specific surface area of catalyst 120-170 is double that the observed for the reference K-OMS-2 crypromelane sample. These results suggest that the excellent activity of catalyst 120-170 cannot be only ascribed to its high specific surface area and the reducibility of the samples is mainly responsible for its high activity.

Regarding the effect of the potassium content, the results reported in the literature are apparently contradictories. It has been described that low potassium content improves the acidic properties of the cryptomelane, and it is beneficial for hydrocarbon oxidation reactions [36]. Zhu also considered that potassium promoted the formation of oxygen vacancies on the surface of cryptomelane and improved the ozone removal [24]. On the other hand, our experimental results and those obtained by Sun et al. [25] indicate that high contents of potassium strongly affect the stability of the cryptomelane structure and the oxygen release occurring at higher temperature negatively affects the activity in oxidation reactions. It is obvious that more detailed work must be devoted to fully understand the effect of the potassium on redox properties and the reaction mechanism.

With aiming to unravel the contribution of the oxygen release peak observed at low temperature during the TPD in the case of Group B, sample 100-1550 was further calcined at 500 ºC. TPD profiles of the 100-1550 and the sample calcined at 500 ºC are depicted in Figure 8 (left). We can observe that calcination strongly reduces the contribution of the first $O_2$ releasing peak. This calcination treatment does not affect the intensity of the other two observed peaks, although the positions of the peaks slightly shift to higher temperatures. In fact, the second peak is observed at 546 ºC in the sample of 100-1550, while the oxygen release occurred at 589 ºC in the sample calcined at 500 ºC. A similar shift of 22 ºC is observed for the third peak, as well. Figure 8 (right) shows the catalytic activity of the initial sample (100-1550) and after its calcination at 500 ºC. The activity dramatically decreases after the further calcination. In fact, the sample after calcination at 500 ºC is not active below 100 ºC, while the initial sample reaches 38 % at that temperature. Therefore, it is obvious the enormous contribution to the catalytic performance of the oxygen associated to the first peak observed in the TPD. It should be remembered that this peak is commonly attributed to desorption of surface oxygen species and/or labile surface lattice oxygen that normally are considered as beneficial for the oxidation reactions. It must also be mentioned that the surface area of the sample after calcination at 500 ºC decreased to 70 m$^2$/g, indicating that effectively the oxygen of the first peak can be directly related with the oxygen located on the surface of the catalyst. We previously discussed that it is commonly accepted that the reducibility of the sample strongly determines the catalytic properties. This result supports the previous work of Santos et al. [21] that an excellent correlation between the position of a shoulder was observed at low temperature in H$_2$-TPR experiments and the catalytic performances expressed as T$_{50}$ (ºC) and T$_{90}$ (ºC). Those species are not strongly stabilized within the oxide lattice, and can be considered as the surface reactive species responsible of the high catalytic activity.

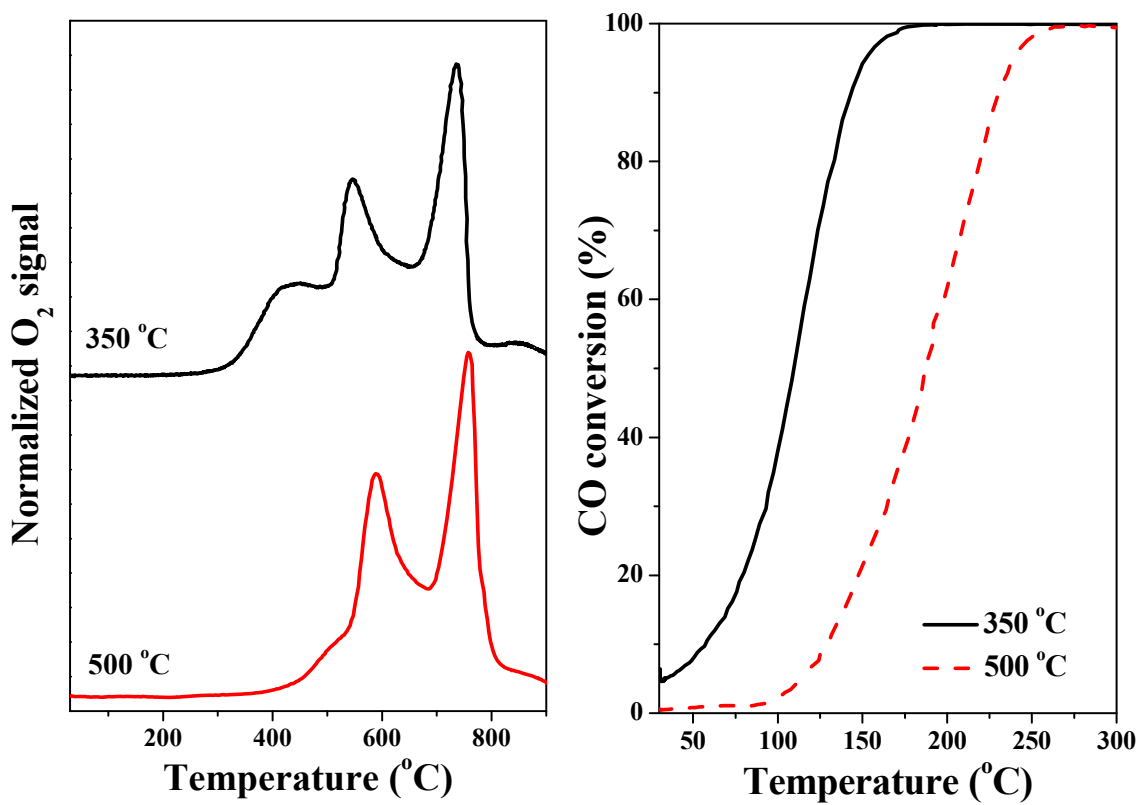

**Figure 8**. Oxygen release during TPD experiment of sample 100-1550 calcined at 350 ºC and 500 ºC (left) and CO oxidation of the same samples (right). The calcination temperature of the sample is included in the plot.

In order to study the catalytic stability of the cryptomelane sample, the catalyst 100-960 in Group B was selected to perform 2 runs of CO oxidation using a 20% oxygen concentration to simulate the air composition. As shown in Figure S9, the catalytic activity is slightly improved by increasing the oxygen concentration from 0.6% to 20%. On the other hand, the activity of the sample in the second cycle remains nearly the same level and only a marginal deactivation is observed below 100 °C. This result demonstrates that the sample exhibits a good stability of the sample under high oxygen concentration.

## 4. Conclusions

Cryptomelane with different potassium concentration can easily be prepared by a novel one-pot hydrothermal synthesis using a potassium permanganate aqueous solution as precursor and ethanol as reducing agent. According to our results, when low ethanol concentration (50 mM) is used in the synthesis, mainly cryptomelane nano-oxide with relatively high specific surface area (100−135 m$^2$/g) and a high potassium content (~16%) is obtained. By increasing the ethanol concentration, the potassium content decreases (~11%) and its surface area increases (160−200 m$^2$/g).

The samples with low content of potassium show excellent conversions of CO to $CO_2$ especially at low temperatures. The superior activity of these samples can be ascribed to the outstanding reducibility of the sample at low temperature that has been highlighted by the TPD and CO-TPR experiments. The lower activity of the samples with high potassium concentration (16%) can be explained as considering the beneficial effect of K on the structural stability of cryptomelane in detriment of labile oxygen on cryptomelane surface. In fact, the TPD results show that those samples are directly reduced to $Mn_3O_4$ at around 795 °C and do not exhibit any oxygen release below this temperature. The higher concentration of potassium on the cryptomelane surface produces a higher concentration of $Mn^{3+}$ that can block the reducibility of the cryptomelane. However, additional characterization is needed for a better understanding of the reduction process at the atomic level.

**Acknowledgement**

This work has been supported by the Ministry of Science and Innovation of Spain/FEDER Program of the EU (MAT2013-50137-EXP, MAT 2013-40823-R and ENE2017-82451-C3-2-R). J. J. Delgado and X. Chen thank the "Ramón y Cajal" Program from MINECO/FEDER of Spain. H. Pan is grateful for financial support to accomplish the PhD study from Chinese Scholarship Council.